\documentstyle[12pt,epsf]{article}
\newcommand{\be}{\begin{equation}}
\newcommand{\ee}{\end{equation}}

\newcommand{\beqa}{\begin{eqnarray}}
\newcommand{\eeqa}{\end{eqnarray}}
\newcommand{\nn}{\nonumber}

\newcommand{\eqref}[1]{(\ref{#1})}

\def\boxit#1{\vbox{\hrule\hbox{\vrule\kern8pt
\vbox{\hbox{\kern8pt}\hbox{\vbox{#1}}\hbox{\kern8pt}}
\kern8pt\vrule}\hrule}}
\def\mathboxit#1{\vbox{\hrule\hbox{\vrule\kern8pt\vbox{\kern8pt
\hbox{$\displaystyle #1$}\kern8pt}\kern8pt\vrule}\hrule}}

\def\IB{\relax\hbox{$\inbar\kern-.3em{\rm B}$}}
\def\IC{\relax\hbox{$\inbar\kern-.3em{\rm C}$}}
\def\ID{\relax\hbox{$\inbar\kern-.3em{\rm D}$}}
\def\IE{\relax\hbox{$\inbar\kern-.3em{\rm E}$}}
\def\IF{\relax\hbox{$\inbar\kern-.3em{\rm F}$}}
\def\IG{\relax\hbox{$\inbar\kern-.3em{\rm G}$}}
\def\IGa{\relax\hbox{${\rm I}\kern-.18em\Gamma$}}
\def\IH{\relax{\rm I\kern-.18em H}}
\def\IK{\relax{\rm I\kern-.18em K}}
\def\IL{\relax{\rm I\kern-.18em L}}
\def\IP{\relax{\rm I\kern-.18em P}}
\def\IR{\relax{\rm I\kern-.18em R}}
\def\IZ{\relax\ifmmode\mathchoice
{\hbox{\cmss Z\kern-.4em Z}}{\hbox{\cmss Z\kern-.4em Z}}
{\lower.9pt\hbox{\cmsss Z\kern-.4em Z}} {\lower1.2pt\hbox{\cmsss
Z\kern-.4em Z}}\else{\cmss Z\kern-.4em Z}\fi}

\def\II{\relax{\rm I\kern-.18em I}}

\def\CA {{\cal A}}

\def\CP {{\cal P}}

\def\CV {{\cal V}}

\begin{document}

\hfill  NRCPS-HE-10-17

\vspace{1cm}
\begin{center}
{\Large ~\\{\it Extension of the Poincar\'e Group
\\
\vspace{0.3cm}
and
\\
\vspace{0.3cm}
Non-Abelian Tensor Gauge Fields
}

}

\vspace{2cm}

{\sl George Savvidy\\
Demokritos National Research Center\\
Institute of Nuclear Physics\\
Ag. Paraskevi, GR-15310 Athens,Greece  \\
}
\end{center}
\vspace{2cm}

\centerline{{\bf Abstract}}

\vspace{12pt}

\noindent
In the recently proposed generalization of the Yang-Mills theory the group of gauge transformation
gets essentially enlarged. This enlargement
involves an elegant mixture of the internal  and   space-time
symmetries. The resulting group is an extension of the Poincar\'e group
with infinitely many generators which carry internal and space-time indices.
This is similar to the super-symmetric extension of the Poincar\'e group,
where instead of an anti-commuting
spinor variable one should introduce a new vector variable.

The construction of irreducible representations of the extended Poincar\'e algebra
identifies a vector variable with the derivative of the Pauli-Lubanski vector over its
length. As a result of this identification the generators of the gauge group
have nonzero components only in the plane transversal to the momentum and are
projecting out non-Abelian tensor gauge fields into the transversal plane,
keeping only their positively definite space-like components.


\newpage

\pagestyle{plain}


\section{\it Introduction}
In the recently proposed generalization of the Yang-Mills theory the group of gauge transformation
gets essentially enlarged \cite{Savvidy:2005fi,Savvidy:2005zm,Savvidy:2005ki}. This enlargement
involves an elegant mixture of the internal  and   space-time
symmetries. The resulting group is an extension of the Poincar\'e group
with infinitely many generators which carry internal and space-time indices.
\cite{Savvidy:2008zy,Savvidy:2010kw}.
This is similar to the super-symmetric extension of the Poincar\'e group
\cite{Golfand:1971iw,Ramond:1971gb,Neveu:1971rx,Volkov:1973ix,Wess:1974tw},
where instead of the anti-commuting
spinor variable  $\theta_{\alpha}$  one should introduce a new commuting vector variable
$e^{\mu}$ \cite{yukawa1,wigner1,Savvidy:2003fx}. Our main concern in this article
is to study the irreducible representation
of this group and its  relation  with the non-Abelian tensor gauge fields.

Let us first overview the gauge transformations which are
defined for the non-Abelian tensor gauge fields and are the source of the proposed
extension of the Poincar\'e group. The  non-Abelian gauge fields are defined as rank-$(s+1)$ tensors
\cite{Savvidy:2005fi,Savvidy:2005zm,Savvidy:2005ki}
$$
A^{a}_{\mu\lambda_1 ... \lambda_{s}}(x),~~~~~s=0,1,2,...
$$
and are totally symmetric with respect to the
indices $  \lambda_1 ... \lambda_{s}  $. The index $a$ numerates the matrix generators $L_a$
of the Lie algebra  $L(G)$. The extended non-Abelian gauge transformations
$\delta_\xi A^{a}_{\mu\lambda_1...\lambda_s}(x)$ of the
tensor gauge fields form a closed group
in which gauge parameters are  totally symmetric tensors $\xi^a_{\lambda_1 ... \lambda_{s}}(x) $
\cite{Savvidy:2005fi,Savvidy:2005zm,Savvidy:2005ki}.
The  extended gauge transformations form a group $\CP$ which is in fact a mixture of the
space-time and internal symmetries \cite{Savvidy:2008zy,Savvidy:2010kw}.
Indeed, the commutator of two such gauge
transformations can be expressed in the form \cite{Savvidy:2005fi,Savvidy:2005zm,Savvidy:2005ki}
$$
[~\delta_{\eta},\delta_{\xi}]~A_{\mu\lambda_1\lambda_2 ...\lambda_s} ~=~
g~ \delta_{\zeta} A_{\mu\lambda_1\lambda_2 ...\lambda_s}~,
$$
where the gauge parameters
 $ \zeta^{a}_{\lambda_1 ... \lambda_{s}}(x)$
are defined by the following equations:
\beqa\label{gaugealgebra}
\zeta^a~~~&=& f^{abc}~ \eta^b  \xi^c \nn\\
\zeta^a_{\lambda_1}~~&=&f^{abc}~ (\eta^b \xi^c_{\lambda_1} + \eta^b_{\lambda_1} \xi^c) \nn\\
\zeta^a_{\lambda_1\lambda_2} &=& f^{abc}~ (\eta^b \xi^c_{\lambda_1\lambda_2}
+ \eta^b_{\lambda_1} \xi^c_{\lambda_2}
+ \eta^b_{\lambda_2} \xi^c_{\lambda_1}  + \eta^b_{\lambda_1\lambda_2} \xi^c), \\
......&.&..........................\nn
\eeqa
This is the algebra of our main concern. Because  the gauge parameters
$\xi^a_{\lambda_1 ... \lambda_{s}}(x)$ have internal
and space-time indices they transform nontrivially under both  groups.
To grasp the structure of this algebra let us consider the first line of the
above equation. It encodes the structure of the internal Lie algebra
$[L_a,L_b] =i f_{abc} L_c$ and we have to ask:
what is the structure of the algebra which is behind the rest of the equations?
Let us introduce for that a translationally invariant commuting vector variable
$e_{\lambda}$ and define
an infinite set of new generators as follows \cite{Savvidy:2005ki}:
\be\label{represocurrentsubalgebra}
L_{a}^{\lambda_1 ... \lambda_{s}} = e^{\lambda_1} ... e^{\lambda_{s}} \otimes L_{a} ~~~~~~~~s=0,1,2......
\ee
These generators carry space-time and internal indices and  transform  under the
operations of both  groups. The algebra of these generators \cite{Savvidy:2005ki}
\be\label{subalgebra1}
[L_{a}^{\lambda_1 ... \lambda_{i}}, L_{b}^{\lambda_{i+1} ... \lambda_{s}}]=if_{abc}
L_{c}^{\lambda_1 ... \lambda_{s} },~~~~s=0,1,2.....
\ee
encodes all equations  (\ref{gaugealgebra}) into a universal one
$
 \zeta(L) =  -i [ \eta(L),  \xi(L)],
$
where the gauge parameters are unified into one function
$
 \zeta(x,L)=
\sum_s {1\over s!} ~\zeta^{a}_{\lambda_1 ... \lambda_{s}}(x) ~L_a^{\lambda_1 ...  \lambda_{s}}.
$

The "current" algebra (\ref{gaugealgebra}) - (\ref{subalgebra1}) is not yet completely defined   because
it does not specify how new generators $L_a^{\lambda_1 ...  \lambda_{s}}$ transform under the
space-time transformations. Taking into
account that the generators $L_a^{\lambda_1 ...  \lambda_{s}}$ are translationally invariant
tensors\footnote{They inherited this property from the vector variable
$e^{\mu}$ in (\ref{represocurrentsubalgebra}).}
of the rank s one can suggest the following extension of
the Poincar\'e algebra \cite{Savvidy:2008zy,Savvidy:2010kw}:
\beqa\label{gaugePoincare}
~&&[P^{\mu},~P^{\nu}]=0,\nn\\
~&&[M^{\mu\nu},~P^{\lambda}] = i(\eta^{\lambda \nu}~P^{\mu}
- \eta^{\lambda \mu }~P^{\nu}) ,\nn\\
~&&[M^{\mu \nu}, ~ M^{\lambda \rho}] = i(\eta^{\mu \rho}~M^{\nu \lambda}
-\eta^{\mu \lambda}~M^{\nu \rho} +
\eta^{\nu \lambda}~M^{\mu \rho}  -
\eta^{\nu \rho}~M^{\mu \lambda} ),\nn\\
~&&[P^{\mu},~L_{a}^{\lambda_1 ... \lambda_{s}}]=0,  \\
~&&[M^{\mu \nu}, ~ L_{a}^{\lambda_1 ... \lambda_{s}}] = i(
\eta^{\lambda_1\nu } L_{a}^{\mu \lambda_2... \lambda_{s}}
-...  -
\eta^{\lambda_s\mu } L_{a}^{\lambda_1... \lambda_{s-1}\nu } ),\nonumber\\
~&&[L_{a}^{\lambda_1 ... \lambda_{i}}, L_{b}^{\lambda_{i+1} ... \lambda_{s}}]=if_{abc}~
L_{c}^{\lambda_1 ... \lambda_{s} }       ~~~(\mu,\nu,\rho,\lambda=0,1,2,3; ~~~~~s=0,1,2,... )\nn
\eeqa
The first three commutators define the Poincar\'e
algebra as its subalgebra. The next two commutators tell that the generators
$L_{a}^{\lambda_1 ... \lambda_{s}}$ are translationally invariant tensors of rank s
and the last commutator defines the "current" subalgebra (\ref{subalgebra1}).
One can check that all Jacoby identities are satisfied and we have an example
of fully consistent algebra, which we shall call an {\it extended Poincar\'e algebra $L(\CP)$
associated with a compact Lie group G.}

As far as the algebra is formulated one can now abstract oneself from the path which led to it and
begin  studing its properties and representations. First of all it is a "gauge invariant"
extension of the Poincar\'e algebra
in a sense that if one  defines a "gauge" transformation of its generators as
$$
L_{a}^{\lambda_1 ... \lambda_{s}} \rightarrow L_{a}^{\lambda_1 ... \lambda_{s}}
+ \sum_{1} P^{\lambda_1}L_{a}^{\lambda_2 ... \lambda_{s}},
$$
then one can check that the algebra $L(\CP)$ remains intact. The second important  property
of the algebra is that it is consistent with the "divergentless" condition imposed
on the generators:
\be\label{divergentless}
P_{\lambda_1} L_{a}^{\lambda_1 ... \lambda_{s}}=0.
\ee
As we shall see, these equations are automatically fulfilled for the matrix representations
which we shall construct in the next sections. This extension of the Poincar\'e algebra
is also consistent with the Coleman-Mandula theorem \cite{Coleman:1967ad,Haag:1974qh}
because it has infinitely many generators and states, which are all massless.

The Casimir operators of the algebra include
$P^2$ and the square of the
Pauli-Lubanski vector $w^{\mu} = {1\over 2 }\epsilon^{\mu\nu\lambda\rho} P_{\nu}~ M_{\lambda\rho}$.
As we shall see, the construction of irreducible representation of the algebra $L(\CP)$ identifies
the translationally  invariant vector $e^{\mu}$  with the derivative of the Pauli-Lubanski vector
over its length, so that  $e^{\mu} = \hat{w}^{\mu}$ (see Appendix A for details) and
the representation takes the form
\be\label{transversalgenerators}
L_{a}^{\bot\lambda_1 ... \lambda_{s}} ~~ =~~  \hat{w}^{\lambda_1}...\hat{w}^{\lambda_s} \otimes L_a~.
\ee
These symmetric generators are transversal
$P_{\lambda_1}~L_{a}^{\bot \lambda_1 ... \lambda_{s}}=0$,
space-like tensors carrying the helicities:
\be
\pm ~s, ~~\pm ~(s-2),~~ \pm~ (s-4),...
\ee
because each vector $\hat{w}^{\lambda_i}$ is a transversal $P_{\mu} \hat{w}^{\mu}=0 $ and
purely spatial unit vector $\hat{ w}^{2} =-1$, which carries a non-zero helicities $h=\pm 1$
\footnote{It is also translationally invariant vector, because
$[P^{\mu},~\hat{w}^{\nu}]=0$.}.

{\it Therefore the generators $L_{a}^{\bot \lambda_1 ... \lambda_{s}}$ are projecting out the components of
the non-Abelian tensor gauge field  $A^{a}_{\mu\lambda_1 ... \lambda_{s}} $ into the plane
transversal to the momentum:
\beqa\label{transversalgaugefields}
~A^{a}_{\mu\lambda_{1}...
\lambda_{s}}  ~L_{a}^{\bot \lambda_1 ... \lambda_{s}} ,\nn
\eeqa
keeping only its positively definite space-like components} of helicities:
\be
\pm (s+1),~~ \begin{array}{c} \pm (s-1)\\ \pm (s-1) \end{array},~~
\begin{array}{c} \pm (s-3)\\ \pm (s-3) \end{array},~~....
\ee
where the lower helicity states have double degeneracy. These are the main results.

We have also found that the current subalgebra (\ref{subalgebra1}) allows "central-like" extension of the form
\beqa\label{deformation}
&&[L_{a}^{\lambda_1}, L_{b}^{\lambda_2}]=if_{abc}
L_{c}^{\lambda_1\lambda_2} -{i\over 4} \delta_{ab}~ \epsilon^{\lambda_1 \lambda_2  \rho \sigma}
w_{\rho} P_{\sigma},\nn\\
&&[L_{a}^{\lambda_1\lambda_2}, L_{b}^{\lambda_3}]=if_{abc}
L_{c}^{(\lambda_1\lambda_2)\lambda_3} +{i\over 4} \delta_{ab}~ (\epsilon^{\lambda_1 \lambda_3 \rho \sigma}
\{w^{\lambda_2} w_{\rho}\} + \epsilon^{\lambda_2 \lambda_3 \rho \sigma}
\{w^{\lambda_1} w_{\rho}\})P_{\sigma},\nn\\
&&[L_{a}^{\lambda_1\lambda_2}, L_{b}^{\lambda_3\lambda_4}]=if_{abc}
L_{c}^{(\lambda_1\lambda_2)(\lambda_3\lambda_4)}
-{i\over 4} \delta_{ab}~ (\epsilon^{\lambda_1 \lambda_3 \rho \sigma}
\{ w^{\lambda_2} \{w_{\rho}  w^{\lambda_4} \} \}
+ \epsilon^{\lambda_2 \lambda_3 \rho \sigma}
\{w^{\lambda_1} \{ w_{\rho} w^{\lambda_4}  \} \} \nn\\
&&~~~~~~~~~~~~~~~~~~~~~~~~~~~~~~~~~~~~~~~~~~~
+\epsilon^{\lambda_1 \lambda_4 \rho \sigma}
\{ w^{\lambda_2} \{w_{\rho}  w^{\lambda_3} \} \}
+ \epsilon^{\lambda_2 \lambda_4 \rho \sigma}
\{w^{\lambda_1} \{ w_{\rho} w^{\lambda_3}  \} \}
)P_{\sigma},\nn\\
&&......................................................................
\eeqa
The deformation (\ref{deformation}) of the current subalgebra (\ref{subalgebra1})
appears  when the vector $e^{\mu} $ is associated with the Pauli-Lubanski vector $e^{\mu} =w^{\mu}$,
which is a {\it noncommutative} vector operator (\ref{noncommutative}).
The derivative of the Pauli-Lubanski vector over its length $\hat{w}^{\nu}$ - is a
{\it commutative} vector operator (\ref{commutingpaulilubanski}) and this is the reason why
the corresponding current subalgebra for the transversal generators
(\ref{transversalgenerators}) is without anomalous terms.
The appearance of the momentum operator in the commutators between  tensor generators
$L_{a}^{\lambda_1 ... \lambda_{s}}$ in this deformed subalgebra is reminiscent of the
super-Poincar\'e algebra (\ref{5}).

In conclusion we shall discuss a possible geometrical interpretation of the group $\CP$
and of the corresponding gauge field theory \cite{Savvidy:2005fi,Savvidy:2005zm,Savvidy:2005ki}.
In the extended Yang-Mills theory, with each space-time point x one should
associate a vector space $\CV_x$ of charged Rarita-Schwinger tensor-spinors fields
$\psi_{\lambda_1 ... \lambda_{s}} ~\hat{w}^{\lambda_1}...\hat{w}^{\lambda_s}$,
on which the action of the group $\CP$
is defined by the group element
$$
U_\xi= \exp{(\sum {1\over s!} ~\xi^{a}_{\lambda_{1}...
\lambda_{s}}L_{a}^{\bot\lambda_1 ... \lambda_{s}}} ).
$$
The extended  gauge
field is a connection which defines a parallel transport of spinor-tensors and
is a $L(\CP)$ algebra valued 1-form. The algebra $L(\CP)$ is not a purely internal algebra - it is
a mixture of {\it internal and space-time  algebras},  which carries not only internal charges,
but also a nonzero helicities. The group $\CP$ acts simultaneously as
a structure group  on the fibers and as an isometry group of the base manifold.

\section{\it Extension of the Poincar\'e Algebra}

As we already explained in the introduction, the gauge fields are defined as rank-$(s+1)$ tensors
\cite{Savvidy:2005fi,Savvidy:2005zm,Savvidy:2005ki} (the Abelian fields are considered in
\cite{schwinger,fierz,fierzpauli,yukawa1,wigner1,Savvidy:2003fx,Weinberg:1964cn,
singh,fronsdal})
$$
A^{a}_{\mu\lambda_1 ... \lambda_{s}}(x),~~~~~s=0,1,2,...
$$
and are totally symmetric with respect to the
indices $  \lambda_1 ... \lambda_{s}  $.  A priory the tensor fields
have no symmetries with
respect to the first index  $\mu$
.The index $a$ numerates the generators $L^a$ of the Lie algebra $L(G)$.

One can think of these tensor fields as appearing in the
expansion of the extended gauge field $\CA_{\mu}(x,L)$ over the generators
$L_{a}^{\lambda_1 ... \lambda_{s}}$
\cite{Savvidy:2005ki}:
\be\label{gaugefield}
{\cal A}_{\mu}(x,L)=\sum_{s=0}^{\infty} {1\over s!} ~A^{a}_{\mu\lambda_{1}...
\lambda_{s}}(x)~L_{a}^{\lambda_1 ... \lambda_{s}}.
\ee
The gauge field $A^{a}_{\mu\lambda_1 ... \lambda_{s}}$ carries
indices $a,\lambda_1, ..., \lambda_{s}$ labeling the generators of {\it the extended
Poincar\'e algebra $L(\CP)$ associated with a compact Lie group G.}
It has infinitely many generators
$L_{a}^{\lambda_1 ... \lambda_{s}} $ and
the corresponding "current" algebra is given by the commutator \cite{Savvidy:2005ki}:
\be\label{subalgebra}
[L_{a}^{\lambda_1 ... \lambda_{i}}, L_{b}^{\lambda_{i+1} ... \lambda_{s}}]=if_{abc}
L_{c}^{\lambda_1 ... \lambda_{s} }.
\ee
Because the generators $L_{a}^{\lambda_1 ... \lambda_{s}}$ are
space-time tensors, they do not commute with the generators of the Poincar\'e algebra $P^{\mu},~M^{\mu\nu}$.
They act on the space-time components of the above generators as
follows  \cite{Savvidy:2008zy,Savvidy:2010kw}:
\beqa\label{gaugePoincare}
~&&[P^{\mu},~P^{\nu}]=0, \label{extensionofpoincarealgebra}\\
~&&[M^{\mu\nu},~P^{\lambda}] = i(\eta^{\lambda \nu}~P^{\mu}
- \eta^{\lambda \mu }~P^{\nu}) ,\nn\\
~&&[M^{\mu \nu}, ~ M^{\lambda \rho}] = i(\eta^{\mu \rho}~M^{\nu \lambda}
-\eta^{\mu \lambda}~M^{\nu \rho} +
\eta^{\nu \lambda}~M^{\mu \rho}  -
\eta^{\nu \rho}~M^{\mu \lambda} ),\nn\\
\nn\\
~&&[P^{\mu},~L_{a}^{\lambda_1 ... \lambda_{s}}]=0,  \\
~&&[M^{\mu \nu}, ~ L_{a}^{\lambda_1 ... \lambda_{s}}] = i(
\eta^{\lambda_1\nu } L_{a}^{\mu \lambda_2... \lambda_{s}}
-\eta^{\lambda_1\mu} L_{a}^{\nu\lambda_2... \lambda_{s}}
+...+
\eta^{\lambda_s\nu } L_{a}^{\lambda_1... \lambda_{s-1}\mu } -
\eta^{\lambda_s\mu } L_{a}^{\lambda_1... \lambda_{s-1}\nu } ),\nonumber\\
\nn\\
\label{alakac}
~&&[L_{a}^{\lambda_1 ... \lambda_{i}}, L_{b}^{\lambda_{i+1} ... \lambda_{s}}]=if_{abc}
L_{c}^{\lambda_1 ... \lambda_{s} }       ~~~(\mu,\nu,\rho,\lambda=0,1,2,3; ~~~~~s=0,1,2,... ),
\eeqa
where the flat space-time metric is
$
\eta^{\mu \nu}= diag(+1,-1,-1,-1).
$
One can check that all Jacoby identities are satisfied and we have an example of a
fully consistent algebra. The commutation relations of the Lorentz generators $ M^{\mu\nu}$ correspond
to the pseoudoorthogonal algebra $SO(1,3)$. The algebra $L(\CP)$ incorporates
the Poincar\'e algebra and an internal algebra $L(G)$ in a nontrivial way, which is different from
the direct product. Specifically the generators $L_{a}^{\lambda_1 ... \lambda_{s}}$
have a nonzero commutation relation with $  M^{\mu\nu} $,
which means that the generators of this new symmetry have nontrivial Lorentz
transformation and that they have a spin different from zero. They
will then relate states of different spins.
The algebra is invariant with respect to the following gauge transformations:\
\beqa
&&L_{a}^{\lambda_1 ... \lambda_{s}} \rightarrow L_{a}^{\lambda_1 ... \lambda_{s}}
+ \sum_{1} P^{\lambda_1}L_{a}^{\lambda_2 ... \lambda_{s}}\nn\\
&&M^{\mu\nu} \rightarrow M^{\mu\nu},~~~~
P^{\lambda} \rightarrow P^{\lambda},
\eeqa
where the sum $\sum_{1}$ is over all inequivalent index permutations.

It is worthy to compare the above extension of the Poincar\'e algebra
with the super-Poincar\'e algebra which is defined as follows
\cite{Golfand:1971iw,Ramond:1971gb,Neveu:1971rx,Volkov:1973ix,Wess:1974tw}:
\beqa\label{superNextensionofpoincarealgebra}
~&&[P^{\mu},~P^{\nu}]=0, \\
~&&[M^{\mu\nu},~P^{\lambda}] = i(\eta^{\lambda \nu}~P^{\mu}
- \eta^{\lambda \mu }~P^{\nu}) ,\nn\\
~&&[M^{\mu \nu}, ~ M^{\lambda \rho}] = i(\eta^{\mu \rho}~M^{\nu \lambda}
-\eta^{\mu \lambda}~M^{\nu \rho} +
\eta^{\nu \lambda}~M^{\mu \rho}  -
\eta^{\nu \rho}~M^{\mu \lambda} ), \nn\\
\nn\\
\label{superNextensionofpoincarealgebra1}
~&&[P^{\mu},~\Theta^{i}_{\alpha}]=0, \\
~&&[M^{\mu \nu}, ~ \Theta^{i}_{\alpha}] = {i\over 2} (
\gamma^{\mu\nu} \Theta^{i})_{\alpha} ,~~~~~~
\gamma^{\mu\nu}={1\over 2} [\gamma^{\mu},\gamma^{\nu}]  \nn\\
\nn\\
\label{5}
~&&\{ \Theta^{i}_{\alpha}, \Theta^{j}_{\beta} \}=-2~\delta^{ij} (\gamma^{\mu} C)_{\alpha\beta}
 P_{\mu},~~~~~i=1,...,N
\eeqa
Both algebras have Poincar\'e algebra (\ref{gaugePoincare}), (\ref{superNextensionofpoincarealgebra})
as subalgebra.  The next two commutators (\ref{extensionofpoincarealgebra})
and (\ref{superNextensionofpoincarealgebra1})
express the fact
that the extended generators $\Theta^{i}_{\alpha}$ and $L_{a}^{\lambda_1 ... \lambda_{s}}$
are translationally invariant operators and carry a nonzero spin.
The last commutators (\ref{alakac}) and
(\ref{5}) are
essentially different in both of the algebras, in super-Poincar\'e algebra the generators
$\Theta^{i}_{\alpha}$ anti-commute to the operator $P^{\mu}$, while in our
case $L_{a}^{\lambda_1 ... \lambda_{s}}$ commute to themselves forming an infinite series
of commutators of "current" subalgebra (\ref{subalgebra}) which cannot be truncated,
so that the index $s$ runs from zero to infinity. We have here an example
of an infinitely dimensional current subalgebra   \cite{Faddeev:1984jp}.

The algebra $L(\CP)$ has a simple representation of the following form:
\beqa\label{represofextenpoincarealgebra}
~&& P^{\mu} = k^{\mu} ,\nn\\
~&& M^{\mu\nu} = i(k^{\mu}~ {\partial\over \partial k_{\nu}}
- k^{\nu }~ {\partial \over \partial k_{\mu}}) + i(e^{\mu}~ {\partial\over \partial e_{\nu}}
- e^{\nu }~ {\partial \over \partial e_{\mu}}),\nn\\
~&& L_{a}^{\lambda_1 ... \lambda_{s}} =e^{\lambda_1}...e^{\lambda_s} \otimes L_a ,
\eeqa
therefore it has at least one nontrivial representation.
This representation appeared, when we were considering the
transformation properties of the non-Abelian tensor gauge fields
\cite{Savvidy:2005fi,Savvidy:2005zm,Savvidy:2005ki}. Our aim now is to
study the matrix representations of this algebra. This will allow to specify
the properties of the vector variable $e^{\mu}$ and to identify it with
the derivative of the Pauli-Lubanski vector over its length - $e^{\mu} = \hat{w}^{\mu}$
(see Appendix A for details).
This vector has all required properties:\\
$\alpha$) it is a commuting vector: $[\hat{ w}^{\mu},~ \hat{w}^{\nu}]=0$,\\
$\beta$)  it is translationally invariant: $[P^{\mu},~ \hat{w}^{\nu}]=0,$\\
$\gamma$) it is a transversal vector: $P^{\mu}~ \hat{ w}_{\mu}=0$,\\
$\delta$) it is a unit space-like vector: $\hat{ w}^{2} =-1$, \\
so that the corresponding generators
\be
L_{a}^{\bot\lambda_1 ... \lambda_{s}}
=  \hat{w}^{\lambda_1}...\hat{w}^{\lambda_s} \otimes L_a
\ee
are
transversal tensors, because $P^{\mu}~ \hat{ w}_{\mu}=0$
and $\hat{ w}^{2} =-1$, projecting out the components of
non-Abelian tensor gauge field into the plane
transversal to the momentum $A^{a}_{\mu\lambda_{1}...
\lambda_{s}}  ~L_{a}^{\bot \lambda_1 ... \lambda_{s}} $,
keeping only their positively definite space-like components.

\section{\it Matrix Representations of Extended Poincar\'e Algebra}

The irreducible representation of the Poincar\'e subalgebra
(\ref{gaugePoincare}), (\ref{superNextensionofpoincarealgebra}) can be found by the  method
of induced representations \cite{wigner}. This method consists of finding a representation
of the subgroup of the Poincar\'e group  and boosting it up to a representation
of the full group. One should choose a given momentum $k^{\mu}=\omega(1,0,0,1)$ which
satisfies $k^2 =0$ and then find the subgroup H which leaves $k^{\mu}$ intact
and find a representation of H on the $\vert k^{\mu} >$ states. One should then induce
this representation of H to the whole Poincar\'e group by boosting the frame
with momentum $k^{\mu}$ to the one with an arbitrary momentum. One can show that the result
is independent of the choice of momentum $k^{\mu}$ \cite{wigner}.

We must find H, the group elements of which leave $k^{\mu}=\omega(1,0,0,1)$ intact.
Under the Lorentz group the action of the element
$$
U_{\theta}= \exp{({i\over 2} \theta_{\mu\nu} M^{\mu\nu})}
$$
creates an infinitesimal transformation $k^{\mu} \rightarrow \theta^{\mu}_{~\nu} ~k^{\nu} +k^{\mu}$.
Hence $k^{\mu}=\omega(1,0,0,1)$ is left invariant provided the parameters obey the relations
\be
\theta_{30} =0,~~~\theta_{10} + \theta_{13}=0,~~~\theta_{20} + \theta_{23}=0,
\ee
hence the Lorentz and $L_{a}^{\lambda_1 ... \lambda_{s}}$ generators in subalgebra
H are
\be
M^{12},~~~\Pi^{'}= -M^{10} + M^{13},~~~\Pi^{''}= -M^{20} + M^{23},~~~L_{a}^{\lambda_1 ... \lambda_{s}}
\ee
with  commutators \cite{wigner}
\be\label{Littlealgebra}
[\Pi^{'},\Pi^{''}]=0,~~~[M^{12},\Pi^{'}]=
i\Pi^{''},~~~[M^{12},\Pi^{''}]= -i\Pi^{'}.\label{little}
\ee
We have to study the commutation relations between $M_{12},~\Pi^{'},~\Pi^{''}$ and
$L_{a}^{\lambda_1 ... \lambda_{s}}$. First of all, the relations
\beqa
 [P^{\mu}, ~ L_{a} ] = 0 ,~~~~
 [M^{\mu \nu}, ~ L_{a} ] = 0
\eeqa
show that the generators of the Poincar\'e group commute with the generators of
the internal algebra L(G),
as they should \cite{Coleman:1967ad}. The next level
commutation relation is
$$
 [M^{\mu \nu}, ~ L_{a}^{\lambda }] = i(
\eta^{\lambda \nu } L_{a}^{\mu  }
-\eta^{\lambda\mu} L_{a}^{\nu })
$$
and it provides the structure of the subgroup H with
the rotation $ M^{12}$:
\beqa\label{Littlealgebra0}
&&[M^{12}, ~ L^{0}_{a} ] = [M^{12}, ~ L^{3}_{a} ]=0 \nn \\
&&[M^{12}, ~ L_{a}^{1 }] = +i L_{a}^{2 }\nn\\
&&[M^{12}, ~ L_{a}^{2 }] = -i L_{a}^{1 },
\eeqa
and with translation operators $\Pi^{'}$ and  $\Pi^{''}$:
\beqa\label{extendedalgebra0}
\begin{array}{ll}
 ~[\Pi^{'}, ~ L^{0}_{a} ] =-i L_{a}^{1 } ~~~&[\Pi^{''}, ~ L^{0}_{a} ] =-i L_{a}^{2 } \\
 ~[\Pi^{'} , ~ L^{3}_{a} ]=-i L_{a}^{1 } ~~~&[\Pi^{''} , ~ L^{3}_{a} ]=-i L_{a}^{2 }    \\
 ~[\Pi^{'}, ~ L_{a}^{1 }] = -i (L^{0}_{a}-L_{a}^{3 }) ~~~&[\Pi^{''}, ~ L_{a}^{1 }] = 0 \\
 ~ [\Pi^{'}, ~ L_{a}^{2 }] = 0 ~~~&[\Pi^{''}, ~ L_{a}^{2 }] = -i (L^{0}_{a}-L_{a}^{3 }).
\end{array}
\eeqa
The equations (\ref{Littlealgebra}),(\ref{Littlealgebra0}) and (\ref{extendedalgebra0})
define the first level generators $L^{\lambda}_{a}$. Our aim is to find the matrix solution
of the equations (\ref{Littlealgebra}),(\ref{Littlealgebra0}) and (\ref{extendedalgebra0}).

\subsection{\it Longitudinal Representation}  If the representation of
the algebra (\ref{Littlealgebra}) has $\varrho =0$ and
$\Pi^{'}=\Pi^{''}=0$, the so called $\bf{O}_s$ representations (see Appendix A),
then from (\ref{extendedalgebra0})
it follows that $L^{1 }_a,~L^{ 2}_a$ are trivial
operators. The only nonzero operators are $M^{12},~L^{0 }_a,~L^{3}_a$:
\beqa
&M^{12}=  \sigma    \otimes 1 ,~~~ L^{0}_{a}=   \sigma    \otimes L_a  ,~~~
 L^{3}_{a}= \sigma  \otimes L_a ,
\eeqa
where $\sigma $ is an integer diagonal matrix and we are using the Kronecker product (see Appendix B) and the
representation (\ref{diagonal}). {\it This representation is purely longitudinal because
the transversal components $L^{1 }_a = L^{ 2}_a =0$ of the generator $L^{\lambda}_{a}$ vanish.}
The next level
operators can be easily constructed by computing commutators between low level generators,
as in (\ref{alakac}), that  gives:
\beqa
& L^{00}_{a}= L^{03}_{a} = L^{33}_{a} =\sigma^2    \otimes \tau_a ,~~~~
L^{01}_{a}= L^{02}_{a} = L^{13}_{a} =L^{11}_{a}= L^{22}_{a} = L^{23}_{a} =0
\eeqa
and so on. This representation is isomorphic to the representation considered  in
the previous section (\ref{represocurrentsubalgebra})
 with the vector variable $e^\lambda$ having only longitudinal
 components $e^\lambda = k^{\lambda}$, so that  the generators take the form
\be\label{longitudinal}
L_{a}^{\vert \vert \lambda_1 ... \lambda_{s}} ~~ = ~~ k^{\lambda_1}...k^{\lambda_s} \otimes L_a
\ee
and are obviously divergentless tensors (\ref{divergentless}).
Our aim is to find out representations for the generators $L_{a}^{\lambda_1 ... \lambda_{s}}$
 which have only transversal components.

\subsection{\it Transversal Representation  }
In order to get different representations which shall have nonzero
transversal components we are going to take the
infinitely dimensional representation $\bf{O}(\varrho)$ of the algebra
(\ref{Littlealgebra}) which has $\varrho \neq 0$ (see Appendix A).
We shall take it in the form of Kronecker product with the
identity matrix in the vector space of representations of the internal algebra G:
\beqa\label{repPoincare}
&&M^{12}= +m_{12}   \otimes 1 \nn\\
&&\Pi^{'}~~ = +\pi^{'}   \otimes 1 \nn\\
&&\Pi^{''}~~ = + \pi^{''}    \otimes 1.
\eeqa
Because $L^{0}_{a}$ and $L^{3}_{a}$ commute with $M^{12}$, we shall take them proportional
to $m^{12}$. The other commutators in (\ref{Littlealgebra0}) tell us that $L^{1}_{a}$
and $L^{2}_{a}$ are proportional to $\pi^{'}$ and $\pi^{''}$.
We shall take them also in the form of Kronecker product  with the same matrix
representation $L_a$ of the internal algebra L(G) in order to get
the following representation of the extended little algebra:
\beqa\label{firstlevelgenerators}
&&L^{1}_{a}= + \pi^{''}    \otimes L_a \nn\\
&&L^{2}_{a}= - \pi^{'}   \otimes L_a \nn\\
&&L^{0}_{a}= +m_{12}    \otimes L_a \nn\\
&&L^{3}_{a}= +m_{12}    \otimes L_a ~.
\eeqa
It is easy to check that it fulfills all the  commutation relations
(\ref{Littlealgebra0}) and (\ref{extendedalgebra0}) and has nonzero transversal
components $L^{1 }_a$ and $L^{ 2}_a$ . One can see also
that divergentless equation (\ref{divergentless}) is satisfied:
\be
k_{\lambda } L^{\lambda }=L^{0}_{a}-L_{a}^{3 }=0,
\ee
so that the vector generator  $L^{\lambda}_{a}$  is orthogonal
to the momentum $k^{\mu}$ and cannot be time-like.

Our next task is to construct generators on the second level.
The second level commutation relation is
$$
 [M^{\mu \nu}, ~ L_{a}^{\lambda_1 \lambda_2}] = i(
\eta^{\lambda_1 \nu } L_{a}^{\mu \lambda_2 }
-\eta^{\lambda_1\mu} L_{a}^{\nu \lambda_2}+
\eta^{\lambda_2 \nu } L_{a}^{\mu \lambda_1 }
-\eta^{\lambda_2\mu} L_{a}^{\nu \lambda_1}),
$$
and it provides the structure of the subgroup with
rotation $ M^{12}$:
\beqa\label{secondlevelcomm1}
&&[M^{12}, ~ L^{00}_{a} ] = [M^{12}, ~ L^{03}_{a} ]=[M^{12}, ~ L^{33}_{a} ]=0 \nn \\
&&[M^{12}, ~ L_{a}^{01 }] = +i L_{a}^{02 }\nn\\
&&[M^{12}, ~ L_{a}^{02 }] = -i L_{a}^{01 }\nn \\
&&[M^{12}, ~ L_{a}^{13 }] = +i L_{a}^{23 }\nn\\
&&[M^{12}, ~ L_{a}^{23 }] = -i L_{a}^{13 }\nn \\
&&[M^{12}, ~ L_{a}^{11 }] = +2 i L_{a}^{12 }\nn\\
&&[M^{12}, ~ L_{a}^{22 }] = -2 i L_{a}^{12 }\nn \\
&&[M^{12}, ~ L_{a}^{12 }] = +i (L_{a}^{22 }-L_{a}^{11 }),
\eeqa
and with translation operators $\Pi^{'}$ and  $\Pi^{''}$:
\beqa\label{secondlevelcomm2}
\begin{array}{ll}
~[\Pi^{'}, ~ L^{00}_{a} ] =-2i L_{a}^{01 },~~&[\Pi^{''}, ~ L^{00}_{a} ] =-2i L_{a}^{02 } \\
~[\Pi^{'}, ~ L_{a}^{01 }] = -i L^{00}_{a}+iL_{a}^{03 }-iL^{11}_{a},~~&[\Pi^{''}, ~ L_{a}^{01 }] =-i L^{12}_{a} \\
~[\Pi^{'}, ~ L_{a}^{02 }] = -i L^{12}_{a} ,~~&[\Pi^{''}, ~ L_{a}^{02 }] = -i L^{00}_{a}+iL_{a}^{03 }-iL^{22}_{a} \\
~[\Pi^{'} , ~ L^{03}_{a} ]=-i L_{a}^{01 } -i L_{a}^{13 },~~&[\Pi^{''} , ~ L^{03}_{a} ]=-i L_{a}^{02 } -i L_{a}^{23 }  \\
~[\Pi^{'}, ~ L_{a}^{11 }] = -2iL^{01}_{a} +2i L^{13}_{a},~~&[\Pi^{''}, ~ L_{a}^{11 }] = 2i L^{23}_{a}-2iL^{02}_{a} \\
~[\Pi^{'}, ~ L_{a}^{12 }] = i L^{23}_{a}-iL^{02}_{a},~~&[\Pi^{''}, ~ L_{a}^{12 }] = i L^{13}_{a}-iL^{01}_{a} \\
~[\Pi^{'}, ~ L_{a}^{13 }] = -i L^{11}_{a}+iL_{a}^{33 }-iL^{03}_{a},~~&[\Pi^{''}, ~ L_{a}^{13 }] = -i L^{12}_{a} \\
~[\Pi^{'}, ~ L_{a}^{22 }] =  0,~~&[\Pi^{''}, ~ L_{a}^{22 }] =  0 \\
~[\Pi^{'}, ~ L_{a}^{23 }] = -i L^{12}_{a},~~&[\Pi^{''}, ~ L_{a}^{23 }] = -i L^{22}_{a}+iL_{a}^{33 }-iL^{03}_{a}   \\
~[\Pi^{'}, ~ L_{a}^{33 }] =-2i L^{13}_{a},~~&[\Pi^{''}, ~ L_{a}^{33 }] =-2 i L^{23}_{a}
\end{array}
\eeqa
The current subalgebra (\ref{subalgebra} ) between generators has the following form:
\beqa\label{intergalcomm}
&&[L_{a}^{0}, L_{b}^{0}]=if_{abc}
L_{c}^{00},~[L_{a}^{0}, L_{b}^{1}]=if_{abc}
L_{c}^{01},~
 [L_{a}^{0}, L_{b}^{2}]=if_{abc}
L_{c}^{02},~
~[L_{a}^{0}, L_{b}^{3}]=if_{abc}
L_{c}^{03}\nn \\
&&[L_{a}^{1}, L_{b}^{1}]=if_{abc},~
L_{c}^{11},~ [L_{a}^{1}, L_{b}^{2}]=if_{abc}
L_{c}^{12},~ [L_{a}^{1}, L_{b}^{3}]=if_{abc} L_{c}^{13}\nn \\
&&[L_{a}^{2}, L_{b}^{2}]=if_{abc}
L_{c}^{22},~[L_{a}^{2}, L_{b}^{3}]=if_{abc}
L_{c}^{23},~\nn\\
&&[L_{a}^{3}, L_{b}^{3}]=if_{abc}
L_{c}^{33}~.
\eeqa
Calculating the commutators (\ref{intergalcomm}) we can find the representation for the second level
generators:
\beqa\label{secondlevelgenerators}
&&L^{11}_{a}= + \pi^{''}\pi^{''}     \otimes L_a \nn\\
&&L^{12}_{a}= - \pi^{'} \pi^{''}   \otimes L_a \nn\\
&&L^{22}_{a}= +\pi^{'} \pi^{'}     \otimes L_a \nn\\
&&L^{00}_{a}=L^{03}_{a}=L_{a}^{33 }=  m_{12} m_{12}   \otimes L_a   \nn\\
&&L^{01}_{a}= +{1\over 2}(m_{12}\pi^{''}+\pi^{''}m_{12})   \otimes L_a   \nn\\
&&L^{02}_{a}= -{1\over 2}(m_{12}\pi^{'}+\pi^{'}m_{12})   \otimes L_a   \nn\\
&&L^{13}_{a}= +{1\over 2}(m_{12}\pi^{''}+\pi^{''}m_{12})     \otimes L_a   \nn\\
&&L^{23}_{a}= -{1\over 2}(m_{12}\pi^{'}+\pi^{'}m_{12})     \otimes L_a .
\eeqa
One can check that all commutators (\ref{secondlevelcomm1}) and  (\ref{secondlevelcomm2})
are satisfied,
but what is surprising is that some of the commutators in the current subalgebra
(\ref{intergalcomm}) are getting additional terms:
\beqa\label{centralextension1}
&&[L_{a}^{0}, L_{b}^{1}]=if_{abc} L_{c}^{01} -{i\over 4} \delta_{ab} \Pi^{'},~
 [L_{a}^{0}, L_{b}^{2}]=if_{abc}
L_{c}^{02}-{i\over 4} \delta_{ab} \Pi^{''}, \nn \\
&&[L_{a}^{1}, L_{b}^{3}]=if_{abc} L_{c}^{13}-{i\over 4}
\delta_{ab} \Pi^{'},~ [L_{a}^{2}, L_{b}^{3}]=if_{abc}
L_{c}^{23}-{i\over 4} \delta_{ab} \Pi^{''}.
\eeqa
The appearance of these "central"-like terms\footnote{In reality it is an operator valued
 anomaly of the current subalgebra (\ref{subalgebra}), but just for
abbreviation we shall call it "central"-like extension or simply deformation of algebra $L(\CP)$,
see also  \cite{Faddeev:1984jp}.} is connected with the fact that the
matrices $m_{12}$ and
$\pi$ in the matrix representation (\ref{firstlevelgenerators})  of the vector
generator $L^{\lambda}_{a}$ are not commuting.
This "central"-type extension of the original algebra $L(\CP)$ is of the main interest for us,
because in this representation of tensor generators $L_{a}^{\lambda_1 ... \lambda_{s}}$
have nontrivial transversal components.
As one can check, the above matrix representation also satisfies the divergentless
equation
\be
k_{\lambda } L^{\lambda\mu }_{a}=k_{\mu } L^{\lambda\mu }_{a}=0,
\ee
or, in components,
$$
L^{00}_{a}=L^{03}_{a}=L_{a}^{33 },~~L^{01}_{a}=L^{31}_{a},~~ L_{a}^{02 } =L^{32}_{a}.
$$
We shall see that also in general
\be
k_{\lambda_1} L^{\lambda_1 ... \lambda_{s} }_{a}=0,
\ee
and therefore tensor generators $L^{\lambda_1 ... \lambda_{s} }_{a}$  are orthogonal
to the momentum $k^{\mu}$ and cannot be time-like.

Deformation of the  current subalgebra (\ref{subalgebra}) appears also at the third level.
Calculating commutators (\ref{subalgebra}) we can find the following sequence of generators:
\beqa\label{thirdlevelgenerators}
&&L^{(11)1}_{a}= + \pi^{''}\pi^{''}\pi^{''}     \otimes L_a \nn\\
&&L^{(11)2}_{a}=L^{(12)1}_{a}= - \pi^{'} \pi^{''} \pi^{''}  \otimes L_a \nn\\
&&L^{(12)2}_{a}=L^{(22)1}_{a}= +\pi^{'} \pi^{'} \pi^{''}  \otimes L_a \nn\\
&&L^{(22)2}_{a}= -\pi^{'} \pi^{'} \pi^{'}     \otimes L_a \nn\\
&&L^{(00)0}_{a}=L^{(00)3}_{a}=L_{a}^{(03)3 }= L_{a}^{(33)3 } =m_{12}^3
\otimes L_a  \nn\\
&&L^{(00)1}_{a}=L^{(03)1}_{a}=L^{(33)1}_{a}=
+{1\over 2}(m_{12}^2 \pi^{''}+\pi^{''}m_{12}^2)   \otimes L_a   \nn\\
&&L^{(00)2}_{a}=L^{(03)2}_{a}=L^{(33)2}_{a}=
-{1\over 2}(m_{12}^2\pi^{'}+\pi^{'}m_{12}^2)   \otimes L_a   \nn\\
&&L^{(11)0}_{a}=L^{(11)3}_{a} =
+{1\over 2}(m_{12}  \pi^{''2}+\pi^{''2}m_{12} )   \otimes l_a   \nn\\
&&L^{(22)0}_{a}=L^{(22)3}_{a} = +{1\over 2}(m_{12}  \pi^{'2}+\pi^{'2}m_{12} )   \otimes L_a   \nn\\
&&L^{(01)0}_{a}=L^{(01)3}_{a}=L^{(31)0}_{a}=L^{(31)3}_{a}=
+{1\over 4}(m_{12}^2 \pi^{''}+2m_{12} \pi^{''}m_{12}
+ \pi^{''}m_{12}^2)   \otimes L_a   \nn\\
&&L^{(02)0}_{a}=L^{(02)3}_{a}=L^{(32)0}_{a}= L^{(32)3}_{a}=
-{1\over 4}(m_{12}^2\pi^{'}+2m_{12}\pi^{'}m_{12}
+ \pi^{'}m_{12}^2)   \otimes L_a   \nn\\
&&L^{(01)2}_{a}=L^{(31)2}_{a} =
-{1\over 4}(m_{12}\pi^{'}\pi^{''}+  \pi^{'} m_{12} \pi^{''}
+  \pi^{''}m_{12}\pi^{'}   + \pi^{'} \pi^{''}m_{12} )   \otimes L_a   \nn\\
&&L^{(02)1}_{a}=L^{(32)1}_{a} =
 -{1\over 4}(m_{12}\pi^{'}\pi^{''} +   \pi^{'}m_{12}\pi^{''}+
 \pi^{''}m_{12}\pi^{'}+ \pi^{'}\pi^{'' }m_{12})   \otimes L_a   \nn\\
&&L^{(01)1}_{a}=L^{(31)1}_{a} =
 +{1\over 4}(m_{12}\pi^{''2}+2  \pi^{''}m_{12}\pi^{''}
 + \pi^{''2}m_{12})   \otimes L_a   \nn\\
&&L^{(02)2}_{a}=L^{(32)2}_{a} = +{1\over 4}(m_{12}  \pi^{'2}+2  \pi^{'}m_{12}\pi^{'}+
  \pi^{'2}m_{12} )   \otimes L_a  .
\eeqa
What happens is that the generators $L^{\lambda_1 \lambda_2 \lambda_3}_{a}$  are not any more
symmetric over all  indices $\lambda_1 \lambda_2 \lambda_3$, but are symmetric with respect to
the following interchanges of indices: $L^{(\lambda_1 \lambda_2) \lambda_3}_{a}=
L^{(\lambda_2\lambda_1)  \lambda_3}_{a}=L^{\lambda_3(\lambda_1 \lambda_2 )}_{a}$.
The  "central" terms in the current subalgebra (\ref{subalgebra}) have the following form:
\beqa\label{centralextension2}
&&[L_{a}^{00}, L_{b}^{1}]=if_{abc} L_{c}^{(00)1}
-{i\over 4} \delta_{ab} (M^{12}\Pi^{'}+\Pi^{'}M^{12}),\nn\\
&&[L_{a}^{00}, L_{b}^{2}]=if_{abc} L_{c}^{(00)2}
-{i\over 4} \delta_{ab} (M^{12}\Pi^{''}+\Pi^{''}M^{12}),\nn\\
&& [L_{a}^{01}, L_{b}^{0}]=if_{abc}
L_{c}^{(01)0} + {i\over 8} \delta_{ab} (M^{12}\Pi^{'}+\Pi^{'}M^{12}), \nn \\
&&[L_{a}^{02}, L_{b}^{0}]=if_{abc} L_{c}^{(02)0}
+{i\over 8} \delta_{ab} (M^{12}\Pi^{''}+\Pi^{''}M^{12}),\nn\\
&&[L_{a}^{01}, L_{b}^{1}]=if_{abc} L_{c}^{(01)1}
-{i\over 4} \delta_{ab} \Pi^{'} \Pi^{''} ,\nn\\
&&[L_{a}^{02}, L_{b}^{2}]=if_{abc} L_{c}^{(02)2}
+{i\over 4} \delta_{ab} \Pi^{'} \Pi^{''} ,\nn\\
&&[L_{a}^{01}, L_{b}^{2}]=if_{abc} L_{c}^{(01)2}
-{i\over 4} \delta_{ab}  \Pi^{''2} ,\nn\\
&&[L_{a}^{02}, L_{b}^{1}]=if_{abc} L_{c}^{(02)1}
+{i\over 4} \delta_{ab}  \Pi^{'2} ,\nn\\
&&[L_{a}^{11}, L_{b}^{0}]=if_{abc} L_{c}^{(11)0}
+{i\over 2} \delta_{ab} \Pi^{'} \Pi^{'' } ,\nn\\
&&[L_{a}^{22}, L_{b}^{0}]=if_{abc} L_{c}^{(22)0}
-{i\over 2} \delta_{ab} \Pi^{'} \Pi^{'' } .
\eeqa
In order to construct high rank generators one can proceed in the same way by computing the commutators
between the lower rank generators.
But already at this stage the general structure of the representation under consideration
becomes apparent and one can formulate it in a covariant form using the Pauli-Lubanski
vector $w^{\mu}$. This vector is a non-commutative vector and, as we shall see, the
above deformation (\ref{centralextension1}), (\ref{centralextension2}) of the current subalgebra (\ref{subalgebra}) appears
because of the non-commutativity of this vector. As a next step we shall introduce the vector
$\hat{w}^{\mu}$ which is a derivative of the $w^{\mu}$ over its length and is a
commutative vector. The corresponding representation of the current subalgebra
will be symmetric, transversal and without central-like terms\footnote{Otherwise the term
$[\CA_{\mu}, \CA_{\nu}]$ in the field strength tensor will have anomalous terms.} - the properties of main
importance in our formulation of generalization of the Yang-Mills theory.

\section{\it Covariant Form of Transversal Representation}

Our aim is therefore to find the covariant formulas for the above representation. The crucial
observation is that the Pauli-Lubanski vector
$w^{\mu} = {1\over 2 }\epsilon^{\mu\nu\lambda\rho} P_{\nu}~ M_{\lambda\rho}=
{1\over 2 }\epsilon^{\mu\nu\lambda\rho}M_{\nu \lambda} ~P_{\rho} $
has the following properties:
\be\label{noncommutative}
[P^{\mu},~w^{\nu}]=0, ~~~P^{\mu} w_{\mu}=0, ~~~[w^{\mu},~w^{\nu}]=
-i \epsilon^{\mu\nu\lambda\rho}~ w_{\lambda} P_{\rho},
\ee
that is, it is translationally invariant, transversal and non-commutative vector.
In components it has the following form (see Appendix A):
\be\label{componentsPauliLubanski}
w^{\mu}= (-m_{12}, - \pi^{''}, ~\pi^{'},-m_{12}  ).
\ee
This allows to express in a covariant form the first, the second and the third level
generators (\ref{firstlevelgenerators}),
(\ref{secondlevelgenerators}) and (\ref{thirdlevelgenerators}),
which we found in the previous section,  as follows:
\beqa\label{generators1}
&&L^{\lambda_1}_{a}= -w^{\lambda_1} \otimes L_{a}, ~~~\nn\\
&&L^{\lambda_1 \lambda_2}_{a}=  {1\over 2}(w^{\lambda_1} w^{\lambda_2}
+ w^{\lambda_2}w^{\lambda_1}) \otimes L_{a}, ~~~\nn\\
&&L^{(\lambda_1 \lambda_2)\lambda_3}_{a}=
{1\over 4}(w^{\lambda_1} w^{\lambda_2}w^{\lambda_3}
+ w^{\lambda_2}w^{\lambda_1}w^{\lambda_3}
+ w^{\lambda_3}w^{\lambda_1}w^{\lambda_2}
+ w^{\lambda_3}w^{\lambda_2}w^{\lambda_1}) \otimes L_{a}, ~~~
\eeqa
or defining the anti-commutator
$$
\{A,B\} \equiv {1\over 2} (A B + B A)
$$
we can express the above representation in a compact form and also find the next level
generators:
\beqa\label{generators2}
&&L_{a}= 1 \otimes L_{a}, ~~~\nn\\
&&L^{\lambda_1}_{a}= -w^{\lambda_1} \otimes L_{a}, ~~~\nn\\
&&L^{\lambda_1 \lambda_2}_{a}=   \{w^{\lambda_1} w^{\lambda_2}\} \otimes L_{a}, ~~~\nn\\
&&L^{ (\lambda_1 \lambda_2)\lambda_3 }_{a}=-
\{ \{w^{\lambda_1} w^{\lambda_2} \} ~w^{\lambda_3} \} \otimes L_{a},\nn\\
&&L^{ (\lambda_1 \lambda_2)(\lambda_3 \lambda_4 )}_{a}=
\{ \{w^{\lambda_1} w^{\lambda_2} \} ~\{ w^{\lambda_3} w^{\lambda_4} \} \} \otimes L_{a},\nn\\
&&L^{ ((\lambda_1 \lambda_2)\lambda_3) \lambda_4 }_{a}=
\{ \{  \{w^{\lambda_1} w^{\lambda_2} \}~ w^{\lambda_3} \}~ w^{\lambda_4}  \} \otimes L_{a},\nn\\
&&........................................................................
\eeqa
This actually means that one should associate the vector variable $e^{\mu}$ with the
Pauli-Lubanski vector $e^{\mu} \sim w^{\mu} $  and the generators will take the following
general form:
\be\label{generators3}
L_{a}^{\lambda_1 ... \lambda_{s}} ~~ = ~~ \{...\{w^{\lambda_1}...w^{\lambda_s} \} \otimes L_a,
\ee
where one should understand an appropriate ordering of the non-commuting Pauli-Lubanski vectors,
as it is in (\ref{generators2}). Using explicit matrix form of the
representation (\ref{generators2}) we can calculate the commutators of
the current subalgebra (\ref{subalgebra}), so that it will take  the following form
(see Appendix B for the algebraic rules which allow to calculate commutators of the
matrices given in the form of the Kronecker product)
\footnote{For simplicity we take the generators of the Lie algebra L(G) in the
fundamental representation of SU(2), $\{ L_a,L_b \}= {1 \over 4} \delta_{ab}$  .}:
\beqa
&&[L_{a}^{\lambda_1}, L_{b}^{\lambda_2}]=if_{abc}
L_{c}^{\lambda_1\lambda_2} -{i\over 4} \delta_{ab}~ \epsilon^{\lambda_1 \lambda_2  \rho \sigma}
w_{\rho} P_{\sigma},\nn\\
&&[L_{a}^{\lambda_1\lambda_2}, L_{b}^{\lambda_3}]=if_{abc}
L_{c}^{(\lambda_1\lambda_2)\lambda_3} +{i\over 4} \delta_{ab}~ (\epsilon^{\lambda_1 \lambda_3 \rho \sigma}
\{w^{\lambda_2} w_{\rho}\} + \epsilon^{\lambda_2 \lambda_3 \rho \sigma}
\{w^{\lambda_1} w_{\rho}\})P_{\sigma},\nn\\
&&[L_{a}^{\lambda_1\lambda_2}, L_{b}^{\lambda_3\lambda_4}]=if_{abc}
L_{c}^{(\lambda_1\lambda_2)(\lambda_3\lambda_4)}
-{i\over 4} \delta_{ab}~ (\epsilon^{\lambda_1 \lambda_3 \rho \sigma}
\{ w^{\lambda_2} \{w_{\rho}  w^{\lambda_4} \} \}
+ \epsilon^{\lambda_2 \lambda_3 \rho \sigma}
\{w^{\lambda_1} \{ w_{\rho} w^{\lambda_4}  \} \} \nn\\
&&~~~~~~~~~~~~~~~~~~~~~~~~~~~~~~~~~~~~~~~~~~~~~~~
+\epsilon^{\lambda_1 \lambda_4 \rho \sigma}
\{ w^{\lambda_2} \{w_{\rho}  w^{\lambda_3} \} \}
+ \epsilon^{\lambda_2 \lambda_4 \rho \sigma}
\{w^{\lambda_1} \{ w_{\rho} w^{\lambda_3}  \} \}
)P_{\sigma},\nn\\
&&[L_{a}^{(\lambda_1\lambda_2)\lambda_3}, L_{b}^{\lambda_4}]=if_{abc}
L_{c}^{((\lambda_1\lambda_2)\lambda_3)\lambda_4}
      -{i\over 4} \delta_{ab}~ (\epsilon^{\lambda_1 \lambda_4 \rho \sigma}
\{ \{w^{\lambda_2} w_{\rho}\} w^{\lambda_3}\}
+ \epsilon^{\lambda_2 \lambda_4 \rho \sigma}
\{ \{ w^{\lambda_1} w_{\rho}\} w^{\lambda_3} \}\nn\\
&&~~~~~~~~~~~~~~~~~~~~~~~~~~~~~~~~~~~~~~~~~~~~~~~~+ \epsilon^{\lambda_3 \lambda_4 \rho \sigma}
\{ \{ w^{\lambda_1} w^{\lambda_2}\} w_{\rho} \}
)P_{\sigma},\nn\\
&&......................................................................
\eeqa
where, as we already explained above, the source of this deformation of the current
subalgebra (\ref{subalgebra}) is a non-commutativity of the matrices $m_{12}$
and $\pi$. In covariant formulation we can clearly see that the deformation of the current subalgebra
appears  because the Pauli-Lubanski vector $w^{\mu}$ is a noncommutative vector
- its components (\ref{componentsPauliLubanski}) are not commuting (\ref{noncommutative}).
The appearance of the momentum operator in the commutators between  tensor generators
$L_{a}^{\lambda_1 ... \lambda_{s}}$ in the current subalgebra is reminiscent of the super-Poincar\'e algebra.
The "central"-like terms \footnote{These  general formulas for the "central" terms identically
coincide with (\ref{centralextension1}) and (\ref{centralextension2}).} can be evaluated further and, in particular,
\be
 \delta_{ab} ~C^{\lambda_1 \lambda_2}=-{i\over 4} \delta_{ab}~ \epsilon^{\lambda_1 \lambda_2  \rho \sigma}
w_{\rho} P_{\sigma}=-{i\over 2} \delta_{ab}~ (P^2 M^{\lambda_1 \lambda_2}
- M^{\lambda_1\rho}P_{\rho} P^{\lambda_2}
+ M^{\lambda_2\rho}P_{\rho} P^{\lambda_1} ).
\ee
It is interesting to see how the Jacoby identities are satisfied in  this case.
One can see that
\beqa
&0=[[L_{a}^{\lambda_1 ... \lambda_{i}}, L_{b}^{\lambda_{i+1} ... \lambda_{s}}], P^{\mu}] +
[[L_{b}^{\lambda_{i+1} ... \lambda_{s}}, P^{\mu}] , L_{a}^{\lambda_1 ... \lambda_{i}}] +
[[P^{\mu} , L_{a}^{\lambda_1 ... \lambda_{i}} ] , L_{b}^{\lambda_{i+1} ... \lambda_{s}} ]= \nn\\
&~=[[L_{a}^{\lambda_1 ... \lambda_{i}}, L_{b}^{\lambda_{i+1} ... \lambda_{s}}], P^{\mu}],\nn
\eeqa
therefore the central term $C^{\lambda_1 \lambda_2}$ should commute with the momentum operator,
which indeed takes place.

The main question which remains to be answered is if we can retain the
transversal character of the above representation and at the same time avoid a non-commutativity
of the Pauli-Lubanski vector and as consequence the corresponding central-like terms.
This can be achieved if one considers the
derivative of the Pauli-Lubanski
vector  $w^{\mu}$ over its length - the Casimir invariant
$\varrho$ (\ref{jacoby}), (\ref{replittlegroup}).
Both operators $\pi^{'}$ and $\pi^{''}$
are linear functions in $\varrho$ and the operator $m_{12}$ is $\varrho$ - independent
(see Appendix A), therefore the derivative will have only transversal components.
Thus we shall take the vector variable as follows:
$$
e^{\mu} ~~ = ~~\partial_{\varrho} ~w^{\mu}=
(0,- \partial_{\varrho}~\pi^{''}, \partial_{\varrho}~\pi^{'}, 0  )=
(0,-  \hat{\pi}^{''},  \hat{\pi}^{'}, 0  ) ,
$$
we shall denote it as $\hat{w}^{\mu}$ (see Appendix A):
\be\label{commutingpaulilubanski}
e^{\mu} ~~ = ~~\hat{w}^{\mu} = (0,-  \hat{\pi}^{''},  \hat{\pi}^{'}, 0  ).
\ee
This vector has all required properties:\\
$\alpha$) it is a commuting vector: $[\hat{ w}^{\mu},~ \hat{w}^{\nu}]=0$,\\
$\beta$) it is translationally invariant: $[P^{\mu},~ \hat{w}^{\nu}]=0,$\\
$\gamma$) it is a transversal vector: $P^{\mu}~ \hat{ w}_{\mu}=0$,\\
$\delta$) it is a unit space-like vector: $\hat{ w}^{2} =-1$.\\
\\
Thus we shall define the transversal representation as
\be\label{transversalrep}
L_{a}^{\bot \lambda_1 ... \lambda_{s}} ~~ =  ~~  \hat{w}^{\lambda_1}...~
\hat{ w}^{\lambda_s} \otimes L_a  .
\ee
These generators have components only in the plane transversal to the momentum,
the property of the main importance. The generators fulfill the current subalgebra (\ref{subalgebra})
without deformation, because the vector variable $\hat{ w}_{\mu}$ is a commuting
operator.
Having in hand the explicit matrix representations of the extended Poincar\'e algebra $\CP$
(\ref{generators1}), (\ref{generators2}), (\ref{generators3})  and (\ref{transversalrep})  we
can calculate different traces which are essential components of
the Lagrangian in the extended Yang-Mills theory
\cite{Savvidy:2005fi,Savvidy:2005zm,Savvidy:2005ki}.

\section{\it Killing Form}

Using the explicit matrix representation of the generators (\ref{firstlevelgenerators})
we can compute now the traces (see Appendix C).   The matrix of products has the form
\beqa\label{scalarproducts}
<s \vert tr (L^{\mu}_a L^{\nu}_{b})  \vert s>
&=&\delta_{ab}  \left( s^2 ~{ k^{\mu}k^{\nu} \over \omega^2}-
{\varrho^2\over 2}(\eta^{\mu\nu} - {(k^{\mu}\bar{k}^{\nu}+\bar{k}^{\mu}k^{\nu} \over k \bar{k}})\right)\nn\\
\eeqa
where $k^{\mu}=(\omega,0,0,\omega)$, $\bar{k}^{\mu}=(\omega,0,0,-\omega)$.
Having  the explicit form for the second level generators (\ref{secondlevelgenerators})
we can calculate also the traces:
\beqa\label{scalarproductsleveltwo}
<s \vert tr  (L_a L^{\mu\nu}_{b})  \vert s>
 &=&\delta_{ab}  \left( s^2 ~{ k^{\mu}k^{\nu} \over \omega^2}-
{\varrho^2\over 2}(\eta^{\mu\nu} - {(k^{\mu}\bar{k}^{\nu}+\bar{k}^{\mu}k^{\nu} \over k \bar{k}})\right).\nn
\eeqa
One can observe that in this formulas  the term proportional to $\varrho^2$ is multiplied  by
the projector $\bar{\eta}^{\lambda_1\lambda_2}$ into
the two-dimensional plane transversal to the momentum $k^\mu$ \cite{schwinger}:
\be\label{progector}
\bar{\eta}^{\lambda_1\lambda_2} = - \eta^{\lambda_1 \lambda_2} + { k^{\lambda_1}\bar{k}^{\lambda_2}
+\bar{k}^{\lambda_1}k^{\lambda_2} \over k \bar{k}},~~~~~~~
k_{\lambda_1 }\bar{\eta}^{\lambda_1\lambda_2}=
k_{\lambda_2 }\bar{\eta}^{\lambda_1\lambda_2}=0.
\ee
Because the parameter $\varrho$ is an invariant quantity we can differentiate the trace formula over  $\varrho$
in order to get  invariant products:
\beqa
\partial^2_{\varrho} ~<s \vert tr (L^{\lambda_1}_a L^{\lambda_2}_{b})  \vert s>
=\partial^2_{\varrho} ~<s \vert tr  (L_a L^{\lambda_1 \lambda_2 }_{b})  \vert s>~ = ~
\delta_{ab}  ~ \bar{\eta}^{\lambda_1\lambda_2}.
\eeqa
Having in hand the explicit expressions for the generators (\ref{generators1}) and
(\ref{generators2}) we can also calculate the traces for
high order generators:
\beqa
& \partial^4_{\varrho} ~<s \vert tr (L^{\lambda_1 \lambda_2 }_a L^{\lambda_3 \lambda_4}_{b})  \vert s> =
\partial^4_{\varrho}~ ~<s \vert tr (L^{(\lambda_1 \lambda_2 )\lambda_3}_a L^{ \lambda_4}_{b})  \vert s>= \\
\nn\\
&=\partial^4_{\varrho}~ <s \vert tr (L^{(\lambda_1 \lambda_2)(\lambda_3 \lambda_4) }_a L_{b})  \vert s>
= \partial^4_{\varrho} ~<s \vert tr (L^{((\lambda_1 \lambda_2)\lambda_3) \lambda_4 }_a L_{b})  \vert s>
~ =~\nn\\
\nn\\
&~ = 3~ \delta_{ab} ~ \{~
 \bar{\eta}^{\lambda_1\lambda_2} \bar{\eta}^{\lambda_3\lambda_4}+
 \bar{\eta}^{\lambda_1\lambda_3} \bar{\eta}^{\lambda_2\lambda_4}+
 \bar{\eta}^{\lambda_1\lambda_4} \bar{\eta}^{\lambda_2\lambda_3} ~\} . \nn
\eeqa
All generators $L^{ \lambda_1 ....  \lambda_s }_a$ (\ref{generators1}) ,
(\ref{generators2}) and (\ref{generators3}) on a given level $s$
are ordered polynomials of
the non-commutative (\ref{noncommutative}) Pauli-Lubanski vectors (\ref{generators2}).
But when the indices $~ \lambda_1, .... , \lambda_s$ run in the transverse
plane $~ \lambda_1, .... , \lambda_s =1,2$ then the components of this vector
commute with each other\footnote{These are transversal indices, that is, the
components of the generator  which
are transversal to the vector
$k^{\mu}=(\omega,0,0,\omega)$.} $[w^{1}, w^{2}]=[-\pi^{''},\pi^{' }]=0$.  Thus the
transversal components of the generator
are  polynomials in the commuting matrices $\pi^{'}$ and $\pi^{''}$
 and therefore are insensitive to the ordering. As a result all generators
on a given level $s$  have equal transversal components and therefore
all traces in the last formula are equal to each other.
In order to illustrate this general structure of the generators let us present their
explicit form for the lower rank generators constructed above. They are
\beqa\label{alllevelgenerators}
\begin{array}{ll}
 L^{1}_{a}= + \pi^{''}    \otimes L_a  \\
 L^{2}_{a}= - \pi^{'}   \otimes L_a  \\
\end{array}  ~~~
\begin{array}{lll}
 L^{11}_{a}= + \pi^{''}\pi^{''}     \otimes L_a  \\
 L^{12}_{a}= - \pi^{'} \pi^{''}   \otimes L_a  \\
 L^{22}_{a}= +\pi^{'} \pi^{'}     \otimes L_a  \\
 \end{array}  ~~~
 \begin{array}{llll}
 &L^{(11)1}_{a}= &+ \pi^{''}\pi^{''}\pi^{''}     \otimes L_a  \\
 L^{(11)2}_{a}=&L^{(12)1}_{a}= &- \pi^{'} \pi^{''} \pi^{''}  \otimes L_a  \\
 L^{(12)2}_{a}=&L^{(22)1}_{a}= &+\pi^{'} \pi^{'} \pi^{''}  \otimes L_a  \\
 &L^{(22)2}_{a}= &-\pi^{'} \pi^{'} \pi^{'}     \otimes L_a  \\
\end{array}           \nn  .
\eeqa
In general, the transversal components of the generator $L^{ \lambda_1 ....  \lambda_s }_a$
have the following structure:
\be
L^{\overbrace{1...1}^{k} \overbrace{2...2}^{m}  }_a =
( \pi^{''})^k (-\pi^{'})^m ~\otimes L_a ~~,
\ee
where $k+m=s$ and is therefore proportional to $\varrho^s$ (\ref{diagonal1}). The components of the
generator $L^{ \lambda_1 ....  \lambda_s }_a$  (\ref{generators3}) with indices $\lambda_1 ....  \lambda_s$
within which there are one or more
longitudinal coordinates $\lambda=0,3$ will have instead of $\pi$  generators
the $m_{12}$ generator, which does not carry the parameter $\varrho$.
These longitudinal components carry less power  of the parameter $\varrho^h$,
where $h <  s$. This observation has an important consequence, because it follows that the
s-order derivative of the generator $L^{ \lambda_1 ....  \lambda_s }_a$ over $\varrho$
will "filter out" only transversal components. We shall define transversal generators
$L^{\bot \lambda_1 ....  \lambda_s }_a$ as s-order derivative of the generator
$L^{ \lambda_1 ....  \lambda_s }_a$ over $\varrho$:
\be\label{transversalrep1}
L^{\bot ~\lambda_1 ....  \lambda_s }_a = \partial^{s}_{\varrho} ~ L^{ \lambda_1 ....  \lambda_s }_a
= \partial_{\varrho} ~w^{\lambda_1} ...~\partial_{\varrho} ~w^{\lambda_s} \otimes L_a =
\hat{w}^{\lambda_1}...~
\hat{ w}^{\lambda_s} \otimes L_a.
\ee
Thus we have seen that analysis of the traces leads us to the same expression for
purely transversal representation (\ref{repPoincare}),(\ref{transversalrep}) and
(\ref{transversalrep1}) of the  algebra $L(\CP)$.

We can now calculate the product of two transversal generators of the total rank $2n$:
$$
L^{\bot~\lambda_{1}... \lambda_{i} }_a
L^{\bot~\lambda_{i+1}... \lambda_{2n}}_{b}.
$$
Its components are the aggregates of $\hat{\pi}$ matrices, so that
calculating the corresponding traces
we come up with the general formula for the Killing form:
\beqa
& {1 \over (2n-1)!!}~ ~<s \vert tr (L^{\bot~\lambda_{1}... \lambda_{i} }_a
L^{\bot~ \lambda_{i+1}... \lambda_{2n}}_{b})  \vert s> ~ =~   \delta_{ab} ~ \sum_{P}
 \bar{\eta}^{\lambda_{i_1}\lambda_{i_2}}......\bar{\eta}^{\lambda_{i_{2n-1}}\lambda_{i_{2n}}},
\eeqa
which contains a symmetric product of projector (\ref{progector}) \cite{schwinger}.
This traces are an
essential ingredients of the Lagrangian, which is a quadratic form in  field
strength tensor \cite{Savvidy:2005fi,Savvidy:2005zm,Savvidy:2005ki}.

\section{\it Conclusion}

In our studies of a generalization of the Yang-Mills theory it has been
found that the group of gauge transformations gets  essentially enlarged
\cite{Savvidy:2005fi,Savvidy:2005zm,Savvidy:2005ki}. This enlargement
involves an elegant mixture of the internal  and  space-time
symmetries. The resulting group is an extension of the Poincar\'e group
with an infinitely many generators which carry internal and space-time
indices \cite{Savvidy:2008zy,Savvidy:2010kw}. Our main concern in this article
was to study the irreducible representation
of this group and its relation  with the generalization of the Yang-Mills theory.

We have found two essentially different representations of the algebra $L(\CP)$. The first one can be
characterized as  {\it longitudinal representation}, because its tensor generators
$L_{a}^{ \vert \vert \lambda_1 ... \lambda_{s}}$ are built up as symmetric polynomials of momentum (\ref{longitudinal})
$$
L_{a}^{\vert \vert \lambda_1 ... \lambda_{s}} ~~ = ~~ k^{\lambda_1}...k^{\lambda_s} \otimes L_a
$$
and are obviously divergentless: $k_{\lambda_1}~L_{a}^{\vert \vert \lambda_1 ... \lambda_{s}}=0$.
If the gauge fields take value in this representation of the algebra
$\CP$ then they reduce to a collection of the Yang-Mills fields $B^{a}_{\mu}$
of the form
$$
 ~A^{a}_{\mu\lambda_{1}...
\lambda_{s}} ~L_{a}^{\vert \vert \lambda_1 ... \lambda_{s}}  ~,
$$
where each of these tensor fields carries the highest helicity $h= \pm 1$. There are no high spin states in
this representation of $\CP$ and, as a consequence, in the corresponding gauge theory.

The second,  {\it the transversal  representation} of the generators
$L_{a}^{\bot \lambda_1 ... \lambda_{s}}$ is build up as symmetric
polynomials of the vector variable $\hat{w}^{\mu}$ which is a derivative
of the Pauli-Lubanski vector
over its length (\ref{commutingpaulilubanski}), (\ref{transversalrep}):
$$
L_{a}^{\bot \lambda_1 ... \lambda_{s}} ~~ =  ~~  \hat{w}^{\lambda_1}...~
\hat{ w}^{\lambda_s} \otimes L_a.
$$
These symmetric generators are transversal
$k_{\lambda_1}~L_{a}^{\bot \lambda_1 ... \lambda_{s}}=0$,
space-like tensors carrying the helicities:
\be
\pm ~s, ~~\pm ~(s-2),~~ \pm~ (s-4),...
\ee
because each vector $\hat{w}^{\lambda_i}$ is a transversal $k_{\mu} \hat{w}^{\mu}=0 $ and
purely spatial unit vector $\hat{ w}^{2} =-1$, which carries a non-zero helicities $h=\pm 1$.
{\it Therefore the generators $L_{a}^{\bot \lambda_1 ... \lambda_{s}}$ are projecting out the components of
the non-Abelian tensor gauge field  $A^{a}_{\mu\lambda_1 ... \lambda_{s}} $ into the plane
transversal to the momentum:
\beqa\label{transversalgaugefields}
~A^{a}_{\mu\lambda_{1}...
\lambda_{s}}  ~L_{a}^{\bot \lambda_1 ... \lambda_{s}} ,\nn
\eeqa
keeping only its positively definite space-like components} of helicities:
\be
\pm (s+1),~~ \begin{array}{c} \pm (s-1)\\ \pm (s-1) \end{array},~~
\begin{array}{c} \pm (s-3)\\ \pm (s-3) \end{array},~~....
\ee
where the lower helicity states have double degeneracy. This can be seen
in the frame which is associated with the fixed momentum
$k^{\mu}=\omega(1,0,0,1)$, so
that the tensor gauge fields describe transversal, positive norm states of
the highest helicity $s+1$.
Because the whole construction is based
on the representation of the space-time group $\CP$ the above statement is \
relativistically invariant (see also the Appendix A).

In conclusion one can suggest the following geometrical interpretation of the group $\CP$
and of the corresponding gauge field theory \cite{Savvidy:2005fi,Savvidy:2005zm,Savvidy:2005ki}.
In the Yang-Mills theory \cite{yang,chern} with each point of the space-time is associated
a vector space $V_x$ of charged spinor fields $\psi$,
on which the action of the Lie group G is defined as  $\psi^{'}=U_\xi \psi$.
The gauge field $A_{\mu}$ is a connection
which defines a parallel transport of $\psi$: $\delta \psi = i g A_{\mu} \psi dx^{\mu}$
and is a Lie algebra valued 1-form.

In the extended Yang-Mills theory, with each space-time point x one should
associate a vector space $\CV_x$ of charged Rarita-Schwinger tensor-spinors fields
$\psi_{\lambda_1 ... \lambda_{s}} ~\hat{w}^{\lambda_1}...\hat{w}^{\lambda_s}$
\cite{Savvidy:2005zm,rarita,singh1,fronsdal1},
on which the action of the group $\CP$
is given by the group element
$
U_\xi= \exp{(\sum {1\over s!} ~\xi^{a}_{\lambda_{1}...
\lambda_{s}}L_{a}^{\bot\lambda_1 ... \lambda_{s}}} )
$. The extended  gauge
field $\CA(x,L)$ is a connection which defines a parallel transport of spinor-tensors and
is a $L(\CP)$ algebra valued 1-form. The algebra $L(\CP)$ is not a purely internal algebra - it is
a mixture of {\it internal and space-time  algebras},  which carries not only internal charges,
but also a nonzero helicities. The group $\CP$ acts simultaneously as
a structure group  on the fibers and as an isometry group of the base manifold.

I would like to thank Prof.Ludwig Faddeev for stimulating discussions and his interpretation
of the algebra (\ref{gaugealgebra}) as an example of a current algebra.

\section{\it Appendix A. Representations of the Poincer\'e Group}
The Poincar\'e group has an invariant subgroup, consisting of all displacements $U_a
=\exp{(i a_{\mu} P^{\mu})}$, and it is
Abelian group. Since the group of displacements is Abelian, the unitary irreducible representations
are one-dimensional, $U_a \vert k> = \exp (-i k \cdot a) \vert k>$, where $a$ is
the displacement vector and the vector
$\vert k>$ describes a state of the Hilbert space with momentum $k^{\mu}$.
If the representation of the Poincar\'e group contains the representation
$\exp (-i k \cdot a)$ of the displacement group, it also contains all representations
$\exp (-i k^{'} \cdot a)$ of this subgroup with $k^{'} = \theta k $  obtainable from
$k$ by a proper
Lorentz transformation $\theta_{\mu\nu}$. Indeed, since $\vert k>$ describes a state of the Hilbert space
with momentum $k^{\mu}$ it follows that the operation
$U_\theta= \exp{({i\over 2} \theta_{\mu\nu} M^{\mu\nu})}$, which corresponds to the
Lorentz transformation $\theta_{\mu\nu}$, will transform the state $\vert k>$ into the state with momentum
$k^{'} = \theta k $
$$
U_a ~(U_\theta \vert k>) = \exp(-i~\theta k \cdot a)~ U_\theta \vert k>.
$$
It is a consequence of the equations
$
U_a U_\theta = U_\theta U_{\theta^{-1}a}
$ and $~~~k \cdot \theta^{-1} a = \theta k \cdot a$.
Thus if the representation of the Poincar\'e group contains the state $\vert k>$
with momentum $k$, it also contains states with all the momentum
$k^{'} = \theta k$, where $\theta_{\mu\nu}$ is any Lorentz transformation.
Thus, in an irreducible representation, all such states can be obtained from a single
one $ \vert k>$ by applying to it all possible Lorentz transformations
 $U_\theta \vert k>$. The length $k^2$ of the momenta is the same
 for all state vectors which are present in the representation space of an irreducible
 representation.

All Lorentz transformations $\theta_{\mu\nu}$ which leave a fixed null vector
$k^{\mu}=\omega(1,0,0,1)$ invariant $\theta~k =k$
form a subgroup called "little group"
\cite{wigner}.
The group of Lorentz transformations which
leave a null vector $k^{\mu}$ invariant is a
two-dimensional Euclidean group of rotations
$U(\theta)=\exp{(-im_{12}\theta)}$ in the plane
transverse to the vector $\vec{k}=(0,0,1)$
and displacements $U^{'}(\alpha)=\exp{(-i\alpha \pi^{'})}$,
$U^{''}(\beta) =\exp{(-i\beta \pi^{''})}$, which are
induced by Lorentz generators
\be
m_{12}= M_{12},~~~\pi^{'}= M_{10} + M_{13},~~~\pi^{''}= M_{20} + M_{23}.
\ee
They form the Euclidean  algebra $E(2)$
\be\label{littlealgebra}
[\pi^{'},\pi^{''}]=0,~~~[m_{12},\pi^{'}]=
i\pi^{''},~~~[m_{12},\pi^{''}]= -i\pi^{'}.\label{little}
\ee
The irreducible unitary representation of the little group uniquely
defines the representation of the Poincer\'e group and it does not
depend on the arbitrary choice of the fixed null vector
$k^{\mu}=\omega(1,0,0,1)$ \cite{wigner}.

Two Casimir operators of the Poincar\'e group are given by
the operators $P^2$ and $w^2$ - square of the Pauli-Lubanski vector
$2 w^{\mu} = \epsilon^{\mu\nu\lambda\rho} P_{\nu}~ M_{\lambda\rho}=
 \epsilon^{\mu\nu\lambda\rho} M_{\nu \lambda}~ P_{\rho}$
\be
-w^2 = {\pi^{'}}^2 + {\pi^{''}}^2
\ee
where $w^{\mu} $ in components is
$$w^{\mu} = (-m_{12},-   \pi^{''}, ~  \pi^{'}, -m_{12}  ).$$
To describe representations of the little group one can
take $m_{12}$ in a diagonal form with integer  eigenvalues
$s=0,\pm 1,\pm 2,...$ and then from
Lorentz subalgebra (\ref{little}) it follows
that both of the $\Pi$ generators have Jacoby
form \cite{wigner}:
\beqa\label{jacoby}
m_{12} = \left( \begin{array}{l}
  .\\
  -2 \\
  ~~-1 \\
  ~~~~~~~~0\\
  ~~~~~~~~~+1\\
  ~~~~~~~~~~~+2\\
  ~~~~~~~~~~~~~~~.
\end{array} \right),~
\pi^{'} =  \varrho \left( \begin{array}{l}
  .\\
  ~~~0~~~~~~~1/2~~~~~~~~~~~~~~~~~~~~~~~\\
  ~~~1/2~~~~~0~~~~~~1/2~~~~~~~~~~~~~~~~~~~~~~~\\
  ~~~~~~~~~~1/2~~~~~0~~~~~~1/2\\
~~~~~~~~~~~~~~~~~~~~1/2,~~~~~0~~~~~~1/2\\
~~~~~~~~~~~~~~~~~~~~~~~~~~~~~1/2,~~~~~0~~~~~~\\
~~~~~~~~~~~~~~~~~~~~~~~~~~~~~~~~~~~~~~~~~~~~~~~.
\end{array} \right), \nn
\eeqa
\beqa
\pi^{''} = \varrho \left( \begin{array}{l}
  .\\
  ~~~0~~~~~~~i/2~~~~~~~~~~~~~~~~~~~~~~~\\
  ~~-i /2~~~~~0~~~~~~i /2~~~~~~~~~~~~~~~~~~~~~~~\\
  ~~~~~~~~~-i /2~~~~~0~~~~~~i /2\\
~~~~~~~~~~~~~~~~~~~~-i /2,~~~~~0~~~~~~i /2\\
~~~~~~~~~~~~~~~~~~~~~~~~~~~~~-i /2,~~~~~0~~~~~~\\
~~~~~~~~~~~~~~~~~~~~~~~~~~~~~~~~~~~~~~~~~~~~~~~~~~~.
\end{array} \right).
\eeqa
These infinitely dimensional representations can be
characterized by the parameter $\varrho$
which can be assumed to be any positive real number.
One can now compute the Casimir operator for
these representations:
\be\label{replittlegroup}
-w^2 = {\pi^{'}}^2 + {\pi^{''}}^2 = \varrho^2 .
\ee
For our purposes it is convenient to define operators $\hat{\pi}^{'}=\pi^{'} / \varrho $
and $\hat{\pi}^{''}= \pi^{''} / \varrho $, so that
\be\label{unitspacelike}
\hat{\pi}^{'2} + \hat{\pi}^{''2}=1~
\ee
and we shall have
\be
\hat{w}^{\mu}= \partial_{\varrho} ~w^{\mu} =
(0,- \partial_{\varrho}~\pi^{''}, \partial_{\varrho}~\pi^{'}, 0  )=
(0,-  \hat{\pi}^{''},  \hat{\pi}^{'}, 0  ).
\ee
Let us define the state with fixed helicity  $s~~(0,\pm 1,\pm 2,...)$ as~~
\be\label{diagonal}
m_{12}|s> =s |s>.
\ee
It can be represented as an
infinite vector with entry one
in the row s:
$$
|s> = \left( \begin{array}{l}
  .\\
  0 \\
  1\\
  0\\
  0\\
  .
\end{array} \right)
$$
The action of the $\pi^{'}$ generators  on this vector can be found
by using (\ref{jacoby}):
\be\label{diagonal1}
\pi^{'} |s> = {\varrho \over 2}(|s-1> + |s+1>),~~~
\pi^{''} |s> = {i\varrho \over 2}(|s-1> - |s+1>) .
\ee
These relations tell us that under the
Lorentz boosts the state with helicity $|s>$ transforms
into the states with helicities $|s\pm 1> $ with the amplitude
proportional to ${\varrho/ 2}$.

In the case when $\varrho =0$ we shall have invariant pure helicity
states $|s>$ which decouple from each other. These are
massless representations ${\bf 0_s }$ of the Poincar\'e algebra.
When $\varrho \neq 0$ we shall have representations,
in which helicity  takes any integer value
$s~=0,\pm 1,\pm 2,...$.   These are so called massless
"continuous spin representations"  - ${\bf 0_{\varrho}}$
of the Poincar\'e algebra \cite{wigner}.
Let us introduce the coherent state of different helicities
$
|\varphi> = \sum_{s} e^{i s \varphi} |s> .
$
Under rotation $U(\theta)$ in xy-plane it will transform as
$$
U(\theta)|\varphi> = \sum_{s} e^{i s \varphi} U(\theta)|s> =
\sum_{s} e^{i s \varphi} e^{i s \theta}|s> = |\varphi + \theta>
$$
and under Lorentz boosts as
\be
\pi^{'} |\varphi> = \sum_{s} e^{i s \varphi} \Pi^{'}|s> =
\sum_{s} e^{i s \varphi} ~{\varrho\over 2} ~(|s-1> + |s+1>) =
\varrho \cos{\varphi} |\varphi>
\ee
and $\pi^{''} |\varphi> = \varrho \sin{\varphi} |\varphi> $.
Thus we have the following representation of the little algebra (\ref{littlealgebra}):
\be\label{trigonometric}
m_{12}= -i {\partial \over \partial \varphi},~~ \pi^{'} = \varrho \cos{\varphi},~~
\pi^{''} = \varrho \sin{\varphi}.
\ee

\section{\it Appendix B. The Kronecker product}

The Kronecker product, defined by $\otimes$, is an operation on two matrices
of arbitrary size resulting in a block matrix
$$
A \otimes B =\left( \begin{array}{lll}
  a_{11} B&...&a_{1n}B\\
  ... &...&...\\
  a_{n1} B&...&a_{nn}B\\
\end{array} \right).
$$
It has mixed-product property.
For matrices A, B, C and D,  for which the matrix product $AC$ and $BD$ is
defined, one has
$$
(A \otimes B ) (C \otimes D) = A C \otimes B D.
$$
This property allows simple calculation of the commutator:
\beqa\label{kronikerproduct}
 [A \otimes B,~ C \otimes D ] &=& [A, C] \otimes B D ~ + ~ C A  \otimes [B,~D ] ~=~~\nn\\
&=&A C \otimes [B,~ D] ~ + ~ [A,~C ] \otimes  D B   ~=~~\nn\\
&=& [A, C] \otimes {1 \over 2}(B D + DB)  ~ + ~{1 \over 2} (A C + C A)  \otimes [B,~D ]\nn\\
&=& [A, C] \otimes \{  B D \}  ~ + ~\{ A C\}  \otimes [B,~D ],
\eeqa
where
$$
\{A,C\} \equiv {1\over 2} (A C + C A).
$$

\section{\it Appendix C. Calculation of Killing Form}

Using the explicit matrix representation of the generators (\ref{firstlevelgenerators})
and representations (\ref{jacoby}) or (\ref{trigonometric}) we can compute  the traces
\beqa
&<s \vert tr (L^{0}_a L^{0}_{b}) \vert s> = \delta_{ab}  s^2,~~~
<s \vert tr (L^{0}_a L^{3}_{b})\vert s> = \delta_{ab}   s^2,~~~
<s \vert tr (L^{3}_a L^{3}_{b})\vert s> = \delta_{ab}  s^2,~~~\nn\\
\nn\\
&<s \vert tr (L^{1}_a L^{1}_{b}) \vert s>= \delta_{ab}   {\varrho^2\over 2},~~~
<s \vert tr (L^{2}_a L^{2}_{b})\vert s> = \delta_{ab}  {\varrho^2\over 2} .\nn
\eeqa
All other traces are equal to zero. Thus the matrix of products has the following form:
\beqa\label{scalarproducts}
<s \vert tr (L^{\mu}_a L^{\nu}_{b})  \vert s> &=&\delta_{ab}
\left(\begin{array}{cccc}
s^2&0&0&s^2 \\
0&{\varrho^2\over 2}&0&0 \\
0&0&{\varrho^2\over 2}& 0\\
s^2&0&0&s^2 \\
\end{array} \right)\nn\\
&=& \delta_{ab}  \left( s^2 ~{ k^{\mu}k^{\nu} \over \omega^2}-
{\varrho^2\over 2}(\eta^{\mu\nu} - {(k^{\mu}\bar{k}^{\nu}+\bar{k}^{\mu}k^{\nu} \over 2 \omega^2})\right)\nn\\
&=&\delta_{ab}  \left( s^2 ~{ k^{\mu}k^{\nu} \over \omega^2}+
{\varrho^2\over 2}\sum_{\lambda=1,2}e^{\mu}_{\lambda}e^{\nu}_{\lambda}  \right),
\eeqa
where $k^{\mu}=(\omega,0,0,\omega)$, $\bar{k}^{\mu}=(\omega,0,0,-\omega)$ and
$e^{\mu}_1=(0,1,0,0)$, $e^{\mu}_2=(0,0,1,0)$.
Having  the explicit form of the second level generators (\ref{secondlevelgenerators})
we can calculate the other traces as well:
\beqa\label{scalarproductsleveltwo}
<s \vert tr  (L_a L^{\mu\nu}_{b})  \vert s> &=&\delta_{ab}
\left(\begin{array}{cccc}
s^2&0&0&s^2 \\
0&{\varrho^2\over 2}&0&0 \\
0&0&{\varrho^2\over 2}& 0\\
s^2&0&0&s^2 \\
\end{array} \right)\nn.
\eeqa
One can observe that in this formulas  the term proportional to $\varrho^2$ is multiplied  by
the projector $\bar{\eta}^{\lambda_1\lambda_2}$ into
the two-dimensional plane transversal to the momentum $k^\mu$ \cite{schwinger}:
\be
\bar{\eta}^{\lambda_1\lambda_2} = - \eta^{\lambda_1 \lambda_2} + { k^{\lambda_1}\bar{k}^{\lambda_2}
+\bar{k}^{\lambda_1}k^{\lambda_2} \over k \bar{k}},~~~~~~~
k_{\lambda_1 }\bar{\eta}^{\lambda_1\lambda_2}=
k_{\lambda_2 }\bar{\eta}^{\lambda_1\lambda_2}=0.
\ee
Because the parameter $\varrho$ is an invariant quantity we can differentiate the traces  over
the parameter $\varrho$  to get the following important formulas:
\beqa
\partial^2_{\varrho} ~<s \vert tr (L^{\lambda_1}_a L^{\lambda_2}_{b})  \vert s>
=\partial^2_{\varrho} ~<s \vert tr  (L_a L^{\lambda_1 \lambda_2 }_{b})  \vert s>~ = ~
\delta_{ab}  ~ \bar{\eta}^{\lambda_1\lambda_2}.
\eeqa

\section{\it Appendix D. Virasoro-Kac-Moody Algebra }
This algebra is defined by the commutators
\beqa
~&&[T^{n},~T^{m}]= (n-m)~T^{n+m} ,\nn\\
~&&[T^{n},~L^{m}_a] = -m ~L^{n+m}_{a},\nn\\
~&&[L^{n}_a, ~ L^{m}_b] =  if_{abc}~L_{c}^{n+m }
\eeqa
and has a simple representation:
$
T^{n} = -z^n {d \over d z}, ~~~L^{n}_a = L_a z^n~,
$
where $z$ is a complex variable on the two-dimensional "space-time" plane. It
can be considered as an analogue of the proposed extension of the Poincar\'e algebra
(\ref{gaugePoincare}) in a
sense that the new vector variable $e_\lambda$  and the  complex variable $z$ play
a similar role, so that the Virasoro subalgebra can be associated
with the Poincar\'e subalgebra and Kac-Moody subalgebra with the "current"
subalgebra (\ref{subalgebra})\footnote{Different extensions of the Poincar\'e algebra recently have been considered
in \cite{Bonanos:2008ez}.}.

\vfill
\end{document}